\documentclass[aps,pre,twocolumn]{revtex4-1}
\usepackage{graphicx}
\usepackage{amsmath,amsfonts,amssymb}
\usepackage{bm}
\usepackage{color}

\usepackage{xcolor}

\usepackage[normalem]{ulem}

\usepackage{soul}
\begin{document}

\title{Time-dependent properties of interacting active matter: 
dynamical behavior of one-dimensional systems of self-propelled particles}

\author{Lorenzo Caprini}
\affiliation{Scuola di Scienze e Tecnologie, Universit\`a di Camerino,  
Via Madonna delle Carceri, I-62032, Camerino, Italy}
  
%\author{Fabio Cecconi}
%\affiliation{Istituto dei Sistemi Complessi (CNR), Via Taurini 19, I-00185 Roma, Italy}

\author{Umberto Marini Bettolo Marconi}
\affiliation{Scuola di Scienze e Tecnologie, Universit\`a di Camerino,  
Via Madonna delle Carceri, I-62032, Camerino, Italy}

\begin{abstract}
We study an interacting  high-density one-dimensional system of self-propelled particles described by the 
Active Ornstein-Uhlenbeck particle (AOUP) model
where, even in the absence of alignment interactions, velocity and energy domains spontaneously form in analogy with those already observed in two dimensions. 
Such domains are regions where the individual velocities are spatially correlated as a result of the interplay between self-propulsion and interactions. Their typical size is controlled by a characteristic correlation length.
In the present work, we focus on a novel and lesser-known aspect of the model, namely its dynamical behavior.
To this purpose, we investigate theoretically and numerically the time-dependent
velocity autocorrelation and spatio-temporal velocity correlation functions. The study of these correlations provides a measure 
of the average life-time and, thus, the stability in time of the velocity domains.
%Our numerical study is supported by theoretical predictions for the whole set of correlations numerically explored.
\end{abstract}

\maketitle

%%%%%%%%%%%%%%%%%%%%%%%%%%%%%%%%%%%%%%%%%%%%%%%%%%%%%%%%%%%%%%%%%%%%%%%%%
\section{Introduction}
%%%%%%%%%%%%%%%%%%%%%%%%%%%%%%%%%%%%%%%%%%%%%%%%%%%%%%%%%%%%%%%%%%%%%%%%%
 In the last period, there has been an upsurge of interest in the study of systems of self-propelled particles
that convert energy into directed and persistent motion.
Certain biological systems such as run-and-tumble bacteria or crawling cells, as well as non-biological systems such as self-driven colloids or artificial swimmers, commonly referred as active matter, can be described in terms of effective models able to capture their salient features~\cite{bechinger2016active, marchetti2013hydrodynamics, gompper20202020}. 
Active  particles display a very rich phenomenology, such as their accumulation at the boundaries~\cite{lee2013active, elgeti2013wall, wensink2008aggregation, caprini2018active} and near rigid obstacles~\cite{mijalkov2013sorting,  nikola2016active, kumar2019trapping, das2020morphological, das2020aggregation, knevzevic2020capillary} or a kind of
non-equilibrium phase-coexistence, known as Motility Induced Phase Separation (MIPS)~\cite{fily2012athermal, redner2013structure, cates2015motility, gonnella2015motility, ma2020dynamic} occurring even in the absence of attractive ~\cite{cates2013active, palacci2013living, buttinoni2013dynamical, bialke2015active, speck2016collective, solon2018generalized, petrelli2018active, chiarantoni2020work, caporusso2020micro, hauke2020clustering} or depletion interactions~\cite{gotzelmann1998depletion}. 
Self-propelled particles are far-from-equilibrium systems, showing several dynamical anomalies which have not a Brownian counterpart \cite{mandal2019motility, bialke2015negative}. A recently reported phenomenon is the formation of large domains characterised by the tendency of the particles towards a common alignement of their velocities.
This fact is somehow surprising since
the particles have spherical symmetry, interact via central potentials, and are not endowed with an alignment mechanism.
These domains have been observed numerically in Ref.~\cite{caprini2020spontaneous} in two-dimensional systems of repulsive self-propelled disks (Active Brownian Particles) both at moderate packing fraction in the phase-coexistence region
and at large packing fraction in homogeneous active liquid, hexatic, and solid phases~\cite{digregorio2018full},
where
domains with aligned velocities can still be observed~\cite{caprini2020hidden}. 
A non-equilibrium phase activity-density diagram has been introduced to represent both homogeneous and inhomogeneous regimes~\cite{caprini2020hidden} and
the structural properties of the system have been compared with the typical size of the aligned domains.
 The model reproduces several experimental results regarding confluent cell monolayers~\cite{basan2013alignment, sepulveda2013collective, garcia2015physics, henkes2020dense}, whose velocity fields display alignment patterns quite similar to the corresponding predictions of the ABP model. 
Hence,  even such a simple model can account for the phenomenology of active matter systems at high density observed in experiments.
In particular, the minimal ingredients to induce the velocity-alignment are i) the excluded volume interaction and ii)
the  persistent self-propulsion, while the long-range attractive force and/or the Vicsek-like velocity interaction are not strictly necessary.

%%%%%%%%%%%%%%%%%%%%%%%%%%%%%%%%%%%%%%%%%%%%%%%%%%%%%%%%%%%%%%%%%%%%%%%%
%----------------------------FIG.1---------------------------------------
\begin{figure}[t!]
\includegraphics[clip=true,keepaspectratio,width=0.98\columnwidth]
{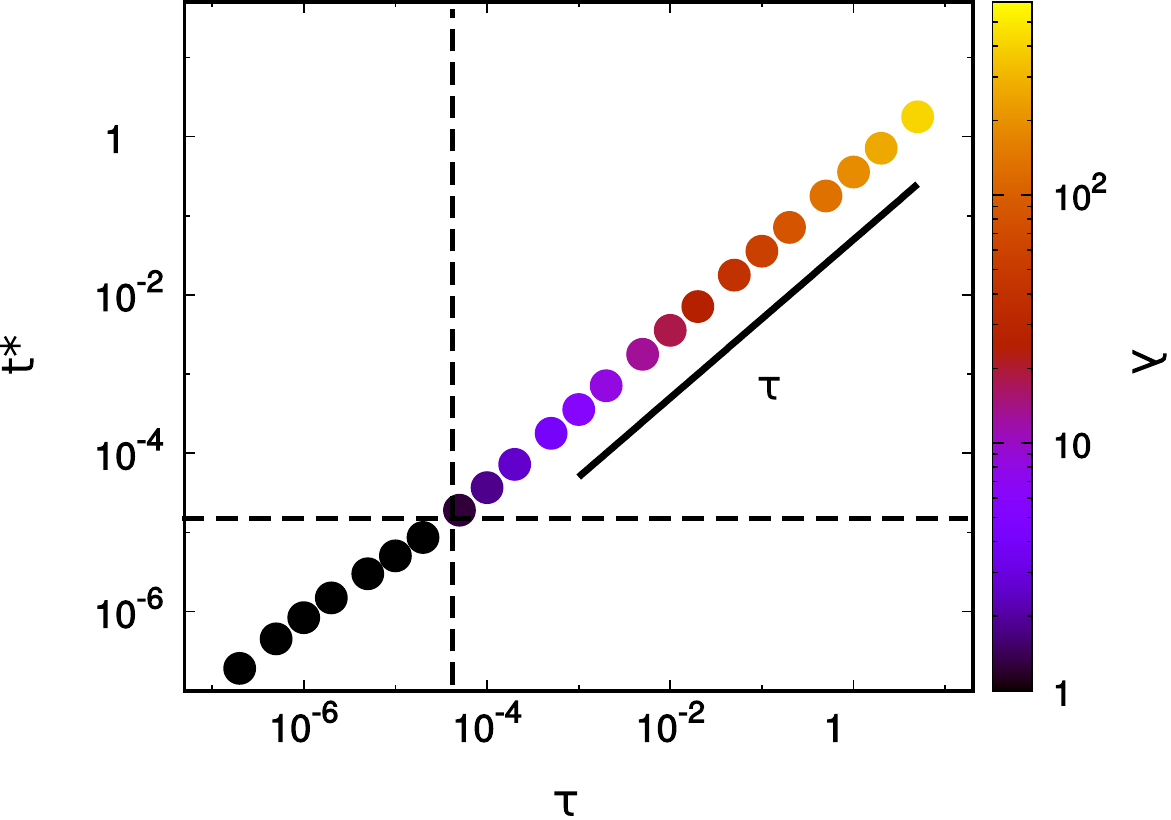}
\caption{\label{fig:lifetimes} 
Life-time, $t^*$, of the aligned domains as a function of $\tau$. Colors represent the size of the aligned domains, in agreement with Fig.~\ref{fig:spatialcorrelation}. Black lines are eye-guides: the solid one shows the linear behavior $\propto \tau$ while the dashed lines are in correspondence of the last value of $\tau$ showing a spatial velocity correlation.
}
\end{figure}
%------------------------------------------------------------------------
At present, notwithstanding the existing detailed information about the velocity domains, obtained
by measuring the equal-time spatial velocity correlations, very little is known about their dynamical properties.
The aim of this work, which will be carried out by numerical and analytical methods, is to characterise how these domains evolve, how stable they are, and what controls their life-time. To achieve this goal, we focus on high-density systems of self-propelled particles in one-dimension.
The motivation of the choice of this low dimensional system is threefold:  a) it considerably reduces the time cost of the numerical simulations, b) it is possible to develop an analytical theory and 
 c) there are situations
where biological swimmers move in highly confining geometries behaving as almost one-dimensional systems.
This is the case of highly confined bacteria~\cite{biondi1998random}, such as E.Coli in microdevices with the size of a single particle, or molecular motors~\cite{chou2011non}. Recently, experimental studies with trains of swimming water-droplets have been performed in microfluidic square channels with sections equal to the droplets' diameter~\cite{illien2020speed, izri2014self}.
In this case, the local velocity alignment between neighbouring droplets has been experimentally observed. 

%%%%
The major outcome of the present investigation is encapsulated in Fig.~\ref{fig:lifetimes}, where the life-time of the velocity domains, $t^*$, defined as the typical decaying-time of the spatio-temporal velocity correlations, is reported as a function of several values of the persistence time, $\tau$ (see Sec.~\ref{Sec:Dynamical} for further details).
Data coloring reflects the value of the correlation length, $\ell$, of the spatial velocity correlations, which quantifies the average size of each velocity domain (see Sec.~\ref{Sec:Dynamical}).
While $\ell \propto \sqrt{\tau}$, our study reveals that $t^*$ displays a linear increase with $\tau$.
The  larger the persistence time of the self-propelled motion, the larger are the typical size and life-time of the velocity domains.

The article is structured as follows: after the introduction of the model reported in Sec.~\ref{Sec:Model}, we ascertain the existence of velocity domains in one-dimensional dense systems of active particles. The steady-state properties, e.g. spatial velocity correlations and correlation lengths, are studied in Sec.~\ref{Sec:velocityDomains}. 
The major novel insight of this work is reported in Sec.~\ref{Sec:Dynamical}, where the velocity autocorrelation and the spatio-temporal velocity correlations are numerically and theoretically investigated.
A final discussion is reported in the conclusive section.
Appendixes contain not only lengthy calculations giving support to the theoretical results of the paper, but also deeper insights into the problem.

%%%%%%%%%%%%%%%%%%%%%%%%%%%%%%%%%%%%%%%%%%%%%%%%%%%%%%%%%%%%%%%%%%%%%%%%
\section{Model}
\label{Sec:Model}
%%%%%%%%%%%%%%%%%%%%%%%%%%%%%%%%%%%%%%%%%%%%%%%%%%%%%%%%%%%%%%%%%%%%%%%%

We study dense systems of $N$ interacting self-propelled particles at density $\rho_0$, employing the Active Ornstein-Uhlenbeck (AOUP) model~\cite{berthier2017active,wittmann2018effective, bonilla2019active, dabelow2019irreversibility, berthier2019glassy, caprini2019comparative, woillez2020active}. The AOUP is a versatile and popular model of active matter
that can reproduce many aspects of the phenomenology of self-propelled particles, including the accumulation near rigid boundaries~\cite{caprini2019activechiral, maggi2015multidimensional, wittmann2016active, marconi2017self, das2018confined} or obstacles and  the motility induced phase separation (MIPS)~\cite{fodor2016far}.
To the best of our knowledge, AOUP is also one of the simplest ways to model self-propelled particles moving in one-dimension in the presence of mutual interactions and/or external forces~\cite{szamel2014self, caprini2018activeescape, fily2019self}. \\ 
In this paper, particles are constrained to move on a line of length $L$ and are subject to periodic boundary conditions.
The particles' positions, $x_i$, evolve with the following stochastic equation:
\begin{equation}\label{eq:motion}
\gamma\dot{x}_i =F_i + \text{f}_i^a \,,
\end{equation}
where $\gamma$ is the drag coefficient and we have neglected the thermal noise due to the solvent since, for many active colloids and bacteria, the thermal diffusivity is usually rather smaller than the effective diffusivity due to the active force \cite{bechinger2016active}.
The term $F_i$ represents the steric interaction between particles and is given by $F_i = - \partial_x U_{tot}$ with $U_{tot}=\sum_{i=1}^{N} U(|x_{i+1} - x_{i}|)$, where $U$ is a truncated and shifted Lennard-Jones (LJ) potential, namely
\begin{equation}
U(r) = 4 \epsilon \left[ \left(\frac{\sigma}{r}\right)^{12}- \left(\frac{\sigma}{r}\right)^6  \right] \,\theta(r-2^{1/6}\sigma)
\label{eq:Ulj}
\end{equation}
where $\theta$ is the Heaviside function and both the energy scale $\epsilon$ and length scales $\sigma$ are set to one, for numerical convenience.
Since the potential chosen has a limited range only the first neighbours mutually interact.
The term $\text{f}_i^a$ models the self-propulsion of the particle $i$ and is described by an Ornstein-Uhlenbeck process according to the  following AOUP dynamics:
\begin{equation}\label{eq:motion2}
\tau\dot{\text{f}}_i^a = -\text{ f}_i^a + \gamma v_0 \sqrt{2 \tau} \xi_i \,,
\end{equation}
where $\xi_i$ is a white noise with zero average and unit variance. 
The parameter $\tau$ is the persistence time and $v_0$ the active velocity associated with the self-propulsion force.
We remark that, in the model employed in this paper, the self-propulsion acts independently on each particle, at variance with 
Vicsek-like models~\cite{ginelli2010large, martin2018collective, sese2018velocity, van2019interrupted}
or more complex dynamics where explicit couplings between velocities and self-propulsions are postulated~\cite{lam2015self, giavazzi2018flocking}.

%{\color{red} Change of variables: the dynamics of the velocity.}\\\\

A convenient method to study the AOUP is achieved by switching from ($x_i$, $\text{f}^a_i$) variables~\cite{marconi2016velocity, fodor2016far, caprini2019activityinduced} to
 position, $x_i$, and velocity, $v_i = \dot{x}_i$ variables.
In one dimension, the equation of motion~\eqref{eq:motion} can be recast as:
\begin{equation}
\label{eq:velocity_dynamics}
\tau\gamma\dot{v}_i = F_i - \gamma\sum_j \Gamma_{ij}(x_i-x_j) v_j + \gamma v_0 \sqrt{2 \tau}  \xi_i \,,
\end{equation}
where the matrix,  $\Gamma_{ij}$,  in general depends on the spatial coordinates of different particles and  contains the derivatives of the potential:
\begin{equation}
\Gamma_{ij} = \delta_{ij} +\frac{\tau}{\gamma} \frac{\partial^2}{\partial x_i\partial x_j} U(x_i-x_j) \,.
\end{equation}
The original active over-damped dynamics is effectively mapped onto a passive under-damped dynamics with a space dependent friction and additional forces which mutually couple the particles' velocities.
When $\tau\gamma \ll 1$ the inertial term can be neglected while for $\tau/\gamma  \,\partial_{x_i}\partial_{x_j} U \ll 1$ uniformly in $x_i$ and $x_j$, the term $\Gamma_{ij}$ reduces to $\delta_{ij}$.
In these cases, Eq.~\eqref{eq:velocity_dynamics} corresponds to an equilibrium over-damped motion 
with long-time effective diffusion coefficient $D=v_0^2 \tau$. 
\subsection{Numerical simulations}
The numerical study presented in this paper has been conducted by setting $v_0=50$ and varying the persistence time of the active force, $\tau$.
Simulations are realised with lengths, $L$, much larger than the persistence length of the active force, in such a way that the condition $L \gg \tau v_0$ is satisfied for the whole range of $\tau$ considered.
Such a regime guarantees that boundary conditions do not play a significant role and that the one-dimensional version of traveling crystals does not occur~\cite{menzel2013traveling, menzel2014active}. 
At variance with previous studies~\cite{locatelli2015active, slowman2016jamming, dolai2020universal, bertrand2018dynamics, barberis2019phase}, we consider high-density regimes, in such a way that the one-dimensional system of active particles is compact enough and neither defects in the periodic arrangement of the particles nor clusters can easily form: the system attains homogeneous configurations for the whole set of parameters numerically explored in this work.

\subsection{The one-dimensional harmonic active crystal}
To develop a suitable analytical theory and interpret the numerical findings, we have considered an 
approximate treatment of \eqref{eq:velocity_dynamics}, by replacing the full LJ potential by its Taylor expansion truncated at the second order:  
\begin{equation}
U_{hc}=U''(\bar x) \sum_{i}^N (x_{i+1}-x_i-\bar{x})^2,
\end{equation}
where the length ${\bar x}=L/N$ is the average inter-particle separation.
Such an approximation works quite well thanks to two conditions: the 
large packing regime and the limited range of the LJ interaction. 
In the simulations, the formation of defects
is practically absent and this makes the mapping onto the active harmonic crystal a successful strategy.  
Being the Langevin equation
relative to the active crystal  \eqref{eq:velocity_dynamics}
linear and  diagonalisable via Fourier modes we can determine all the stationary one-time and two-time correlation functions within this approximation.

\section{Formation of velocity domains: spatial velocity correlations}\label{Sec:velocityDomains}

%----------------------------FIG.2---------------------------------------
\begin{figure}[t!]
\includegraphics[clip=true,keepaspectratio,width=0.98\columnwidth]
{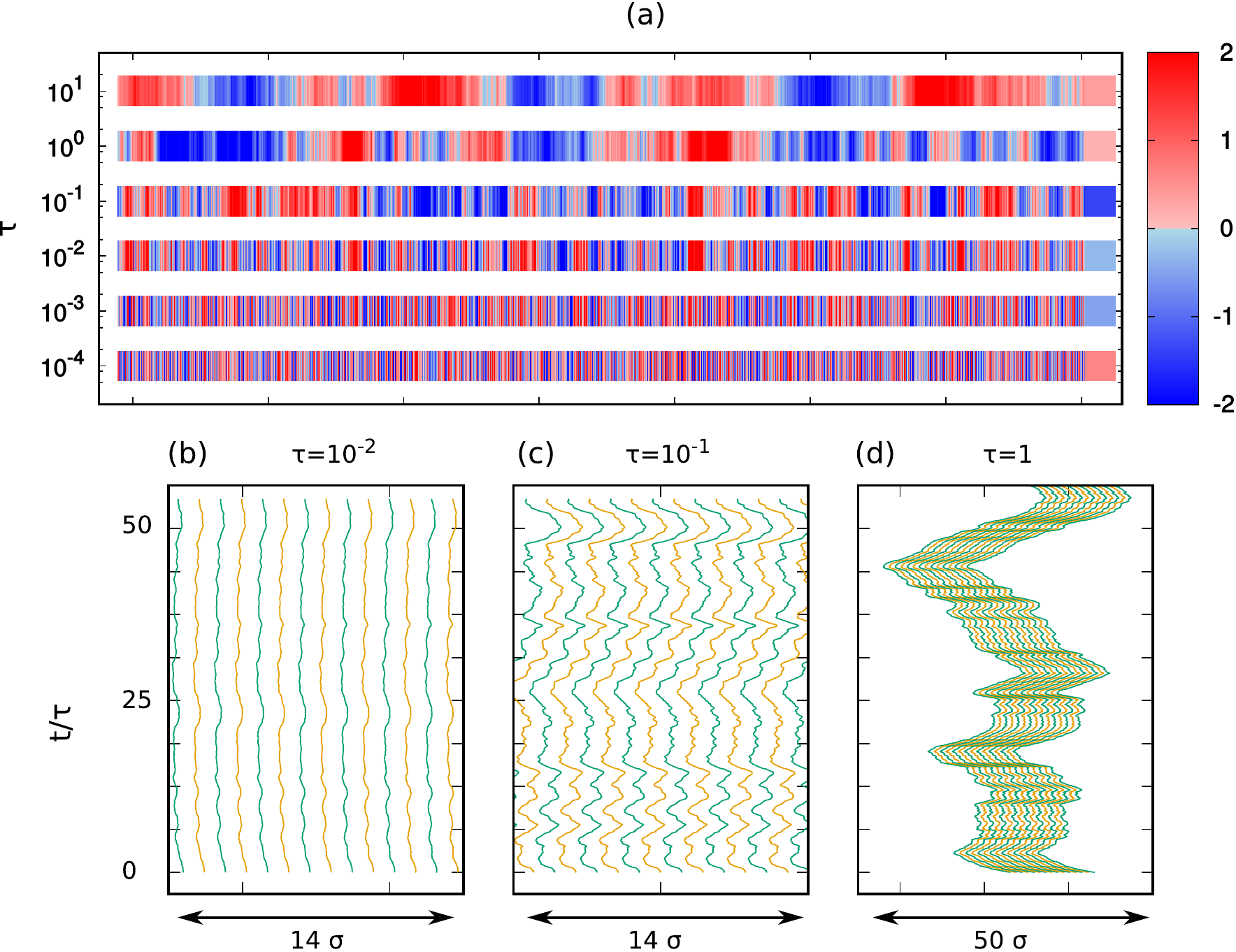}
\caption{\label{fig:snap} 
Panel (a): snapshot configurations for different values of $\tau$. Each configuration is reported along the $x$-axis  of length $L$. For presentation reasons, each particle is represented as a vertical segment.
Configurations with different $\tau$ are reported on the $y$-axis, as shown in the graph, while colors represent the value of the velocity normalised by $\sqrt{\langle v^2\rangle}$.
Panels (b), (c) and (d) are time-trajectories, showing $t/\tau$ vs position of each of 14 neighbouring particles. 
In particular, panel (b), (c) and (d) correspond to $\tau=10^{-2}, 10^{-1}, 1$, respectively.
Neighbouring particles have been drawn with different colors just for presentation reasons.
Simulations are realised with $v_0=50$ and the interaction is given by Eq.~\eqref{eq:Ulj}.
}
\end{figure}
%------------------------------------------------------------------------

Since the 1D AOUP shares the same physics as its 2D companion, also in this case, at high-density 
we expect the formation of spatial domains where the velocities statistically point towards a common direction and have the same modulus.
This is pictorially shown in Fig.~\ref{fig:snap}~(a) where several
one-dimensional instantaneous configurations
of the system are reported for different values of $\tau$.
Particles are represented by vertical segments and are colored according to their velocities, in such a way that a spatial region with the same color can be identified with a velocity domain. 
When $\tau$ increases, the size of the domains grows, as revealed by the color gradients in Fig.~\ref{fig:snap}~(a). 

To understand the formation of the velocity domains, we note that the dynamics \eqref{eq:velocity_dynamics} can be approximated with
\begin{equation}
\label{eq:vicsek_dynamics}
\tau\gamma\dot{v}_i = - \gamma v_i+ \gamma v_0 \sqrt{2 \tau}\xi_i + F_i -2\tau U''(\bar{x}) \left( v_i - w_i \right) \,.
\end{equation}
The additional velocity term, $w_i$, is the mean velocity of the adjacent particles:
$$
w_i = \frac{v_{i+1}+v_{i-1}}{2} \,.
$$
Further details about the derivation of Eq.~\eqref{eq:vicsek_dynamics} are reported in the Appendix~\ref{SecBrown}.
 The first term is an effective Stokes force while the second is assimilable to thermal noise, therefore it
does not lead to any kind of velocity alignment. The term, $F_i$, represents the collective field which constrains the particles
to form and maintain a lattice structure of periodicity $\bar{x}$.
The last term forces the velocity of particle $i$ to assume values equal to the average velocity of its neighbours:
it resembles an interaction {\it \`a la Vicsek}.
We observe that this effective alignment force, dominant if $\tau U''(\bar{x})/\gamma\gg 1$, is a genuine non-equilibrium effect
and leads to the spontaneous local velocity alignment.
On the other hand, in the opposite regime, i.e. $\tau U''(\bar{x})/\gamma\ll 1$, the alignment force is negligible and neighbouring velocities are uncoupled as in the passive Brownian case.

%----------------------------FIG.3---------------------------------------
\begin{figure}[t!]
\includegraphics[clip=true,keepaspectratio,width=0.75\columnwidth]
{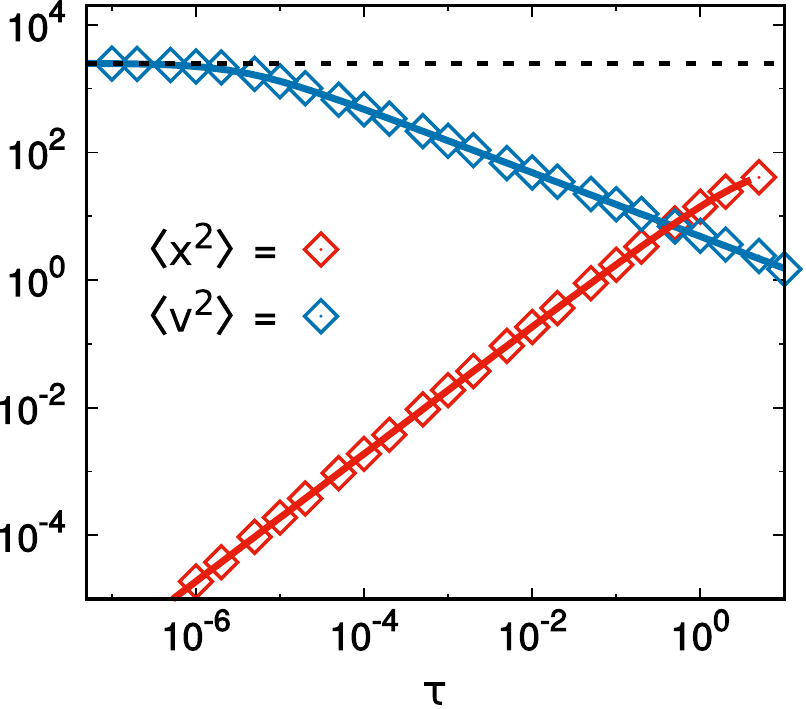}
\caption{\label{fig:Var} 
Variances of position, $\text{Var}(x)=\left\langle \left(x - \langle x \rangle\right)^2\right\rangle$, and velocity, $\langle v^2 \rangle$, as a function of $\tau$.
Colored points are obtained from simulations while colored solid lines from theoretical predictions, Eqs.\eqref{eq:variance_velocity} and \eqref{eq:variance_position} for $\langle v^2 \rangle$ and $\text{Var}(x)$, respectively.
The dashed black line is drawn in correspondence of $v_0^2$, i.e. the asymptotic value of $\langle v^2 \rangle$.
Simulations are realised with $v_0=50$ and the interaction is given by Eq.~\eqref{eq:Ulj}.
}
\end{figure}
%------------------------------------------------------------------------

Let us look at the dynamics from another point of view:  in
Figs.~\ref{fig:snap}~(b)-(d), we sketch the trajectories of a group of adjacent particles: the positions fluctuate around their average value. 
In dense passive systems, the amplitude of these fluctuations is in general much weaker than
in active systems, where the oscillations around the equilibrium positions are rather strong and grow with $\tau$.
While in the small-$\tau$ regime, fluctuations are very small and similar to those of a passive system,
the increase of $\tau$ induces larger fluctuations of a growing number of particles that move cooperatively.
These coherent fluctuations are a direct manifestation of the existence of velocity domains.

The increase of the oscillation amplitude of a single-particle is captured by measuring the variance of the single particle position which is theoretically calculated in the Appendix~\ref{app:variances} in the case of highly dense configurations and almost perfect lattices. Under these assumptions, we obtain:
\begin{equation}
\label{eq:variance_position}
\left\langle \left(x - \langle x \rangle\right)^2\right\rangle = \tau \frac{v_0^2 \gamma}{U''(\bar{x})}\frac{N}{2\pi^2} - \tau^2 \langle v^2\rangle \,,
\end{equation}
where the velocity variance, $\langle v^2 \rangle$, is given by
\begin{equation}
\label{eq:variance_velocity}
\langle v^2 \rangle = \frac{v_0^2}{\left(1+4 \frac{\tau}{\gamma}U''(\bar{x})\right)^{1/2}} \,.
\end{equation}
In the right hand side of Eq. \eqref{eq:variance_position},
the first term coincides with the variance of a passive system  ($\tau=0$) whereas the
 second term gives a negative correction resulting from the formation of correlated velocity domains.

Remarkably, for small values of $\tau$, such that $\tau U''(\bar{x})/ \gamma \ll 1$, the velocity variance is nearly constant and 
is approximately $v_0^2$, while for larger values of $\tau$, $\langle v^2 \rangle$ decreases as $\tau^{1/2}$.
Also an increase of $\rho_0\propto 1/\bar{x} $  makes $U''(\bar{x})$ larger and leads to a decrease of $\langle v^2 \rangle$.
Both observations qualitatively agree with the results relative to 2D interacting particles in non-harmonic potentials where a similar trend was reported~\cite{caprini2019activityinduced}.
Moreover, the variance of the position linearly grows with $\tau$, as shown in Fig.\ref{fig:Var}, in agreement with the qualitative observations of Fig.~\ref{fig:snap}~(b)-(d). 
Indeed, as explained in Appendix~\ref{app:variances}, Eq.\eqref{eq:variance_position} holds 
only for systems with $N \bar x/v_0\tau \gg 1$ and, thus, the growth of the positional variance remains monotonic with $\tau$.
%holds true  in the regime of persistence length smaller than the size of the box, such that $ v_0 \tau  \ll L =N\sigma$.
Interestingly, for $\tau=10^{-1}$ and mostly $\tau=1$, i.e. when $\sqrt{ \left\langle \left(x - \langle x \rangle\right)^2\right\rangle }\gtrsim \bar{x}$, the positional fluctuations are so large as to force particles to synchronise their fluctuations, i.e. to correlate their velocities.

\subsection{Spatial velocity and energy correlations}

%----------------------------FIG.4---------------------------------------
\begin{figure*}[t!]
\includegraphics[clip=true,keepaspectratio,width=0.98\textwidth]
{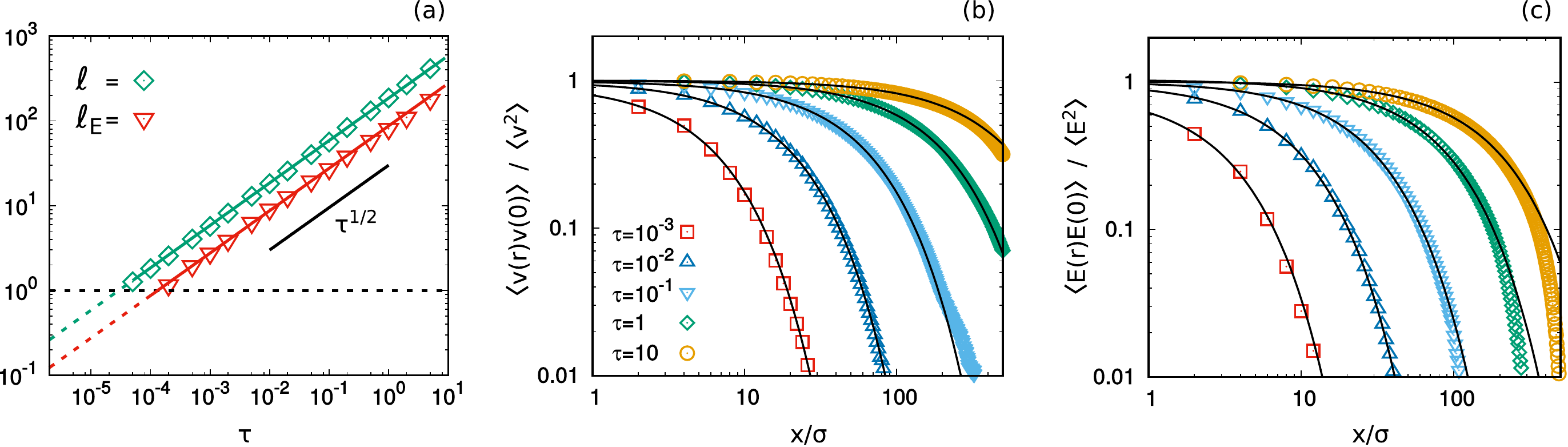}
\caption{\label{fig:spatialcorrelation} 
Spatial correlation functions. Panel (a): $\ell$ and $\ell_E$, i.e. correlation lengths of the velocity and kinetic energy versus $\tau$. Colored data and solid lines are obtained from numerical simulations and Eq.~\eqref{eq:correlation_length} (theoretical prediction), respectively. 
Panels (b) and (c) (sharing the same caption): spatial correlations of the velocities, $\langle v(x) v(0)\rangle/\langle v^2 \rangle$ and energies, $\langle E(x) E(0)\rangle/\langle E^2 \rangle$, respectively, for several values of $\tau$ as shown in the legend. Colored points are obtained by simulations while black solid lines are the theoretical predictions
obtained from Eqs.~\eqref{eq:spatial_corr_theory} and~\eqref{eq:spatialenergy_corr}.
Simulations are realised with $v_0=50$ and the interaction is given by Eq.~\eqref{eq:Ulj}.
}
\end{figure*}
%------------------------------------------------------------------------

To characterise the size of the velocity domains, we study the spatial velocity correlation functions, $\langle v(x) v(0) \rangle$, in the steady-state adapting the strategy of Ref.~\cite{caprini2020hidden} to a one-dimensional system.
In this simple one-dimensional case, $\langle v(x) v(0) \rangle$ can be analytically predicted 
in the active harmonic crystal approximation, as shown in Appendix~\ref{app:SpatialCorrealation}.
We find that, in the stationary regime, the fluctuation amplitude of each velocity mode has the following shape:
\begin{equation}
\label{eq:vvprediction_Fourierspace}
\langle \hat v_q \hat v_{-q}\rangle = \frac{v_0^2}{1+\frac{\tau}{\gamma}\omega_q^2}\,,
\end{equation}
where the frequency $\omega_q$ reads:
\begin{equation}
\label{eq:def_omega2}
\omega_q = \sqrt{2 U''(\bar{x}) \left(1- \cos{\left(q \right)} \right)} \,,
\end{equation}
and $q$ takes values from $-\pi$ to $\pi$ in the limit $N\gg 1$.
Inverting analytically the average~\eqref{eq:vvprediction_Fourierspace} to determine its
real space representation is not so easy without additional approximations, although some formulae
can be found in terms of trascendental functions (see the Appendix~\ref{app:SpatialCorrealation}).
The function represented in Eq.~\eqref{eq:vvprediction_Fourierspace}, being peaked around $q=0$,  is approximated employing a small $q$-expansion. Fourier transforming back to real space, the resulting velocity correlation
displays an exponential spatial decay. In the continuum limit we obtain:
%Adopting a continuum notation for comparison with numerical results, i.e. taking $v_i \to v(x)$, we find:
\begin{equation}
\label{eq:spatial_corr_theory}
\langle v(x) v(0) \rangle\approx  \langle v^2 \rangle\exp{\left(-\frac{x}{\ell}\right)}\,,
\end{equation}
%{\color{blue} Non conviene mettere la denotazione continua? con $v(x)$}
where $\ell$ is the correlation length given by:
\begin{equation}
\label{eq:correlation_length}
\ell= \bar{x}\sqrt{\frac{\tau}{\gamma} U''(\bar{x})}\,.
\end{equation}
 Fig.~\ref{fig:spatialcorrelation}~(b) reports
a set of spatial correlations functions corresponding to different values of $\tau$. It shows the excellent agreement between the numerical data and the  theoretical predictions~\eqref{eq:spatial_corr_theory}.
Based on such an exponential behaviour,  we argue that $\ell$ represents a measure of the size of the velocity domains, since particles within a distance $\approx\ell$ are correlated and, roughly speaking, share similar velocities. 
The average size of a typical velocity domain scales as $\sim \tau^{1/2}$ and increases with $\rho_0$, due to the dependence 
of $\ell$ on the curvature of the potential (higher $\rho_0\propto 1/\bar{x}$ means smaller $\bar{x}$ and, thus, larger $U''(\bar{x})$), as shown by \eqref{eq:correlation_length}.

The spatial  correlation of the kinetic energy, $E(x)=v^2(x)$, shows a similar trend. The two-point average
$\langle E(x) E(0) \rangle$, in the steady state,  is used to measure the size of domains sharing the same kinetic energy.
This observable approaches a non-vanishing asymptotic value, given by $\langle E^2 \rangle^2$.
For this reason, we numerically evaluate the normalised cumulant $\langle E(x) E(0) \rangle_c=[\langle E(x) E(0) \rangle - \langle E^2 \rangle^2]/\langle E^2 \rangle^2$. 
This spatial correlation is analytically calculated in the Appendix~\ref{app:SpatialCorrealation} and shows the following exponential shape:
\begin{equation}
\label{eq:spatialenergy_corr}
\langle E(x) E(0) \rangle_c \approx  \langle E^2 \rangle\exp{\left(-\frac{x}{\ell_E}\right)}\,,  
\end{equation}
where $\ell_E$ is its correlation length which reads:
\begin{equation}
\ell_E= \sqrt{\frac{\tau}{2\gamma} U''(\bar{x})}\,.
\end{equation}
Theoretical predictions fairly agree with data as revealed both in panel (a) and (c) of Fig.~\ref{fig:spatialcorrelation}, reporting $\ell_E$ and the energy correlations, respectively.
Also in this case, $\ell_E$ corresponds to the average size of energy domains, which is smaller than $\ell$ by a factor 
$2^{-1/2}$ and maintains the same scaling with the parameters of the model. 
As a consequence, in the one-dimensional case, the spatial energy correlations do not contain further information concerning the spatial velocity correlations.
\section{Dynamical properties of the velocity domains}\label{Sec:Dynamical}

%%%%%%%%%%%%%%%%%%%%%%%%%%%%%%%%%%%%%%%%%%%%%%%%%%%%%%%%%%%%%%%%%%%%%%%%
%----------------------------FIG.5---------------------------------------
\begin{figure*}[t!]
\includegraphics[clip=true,keepaspectratio,width=0.98\textwidth]
{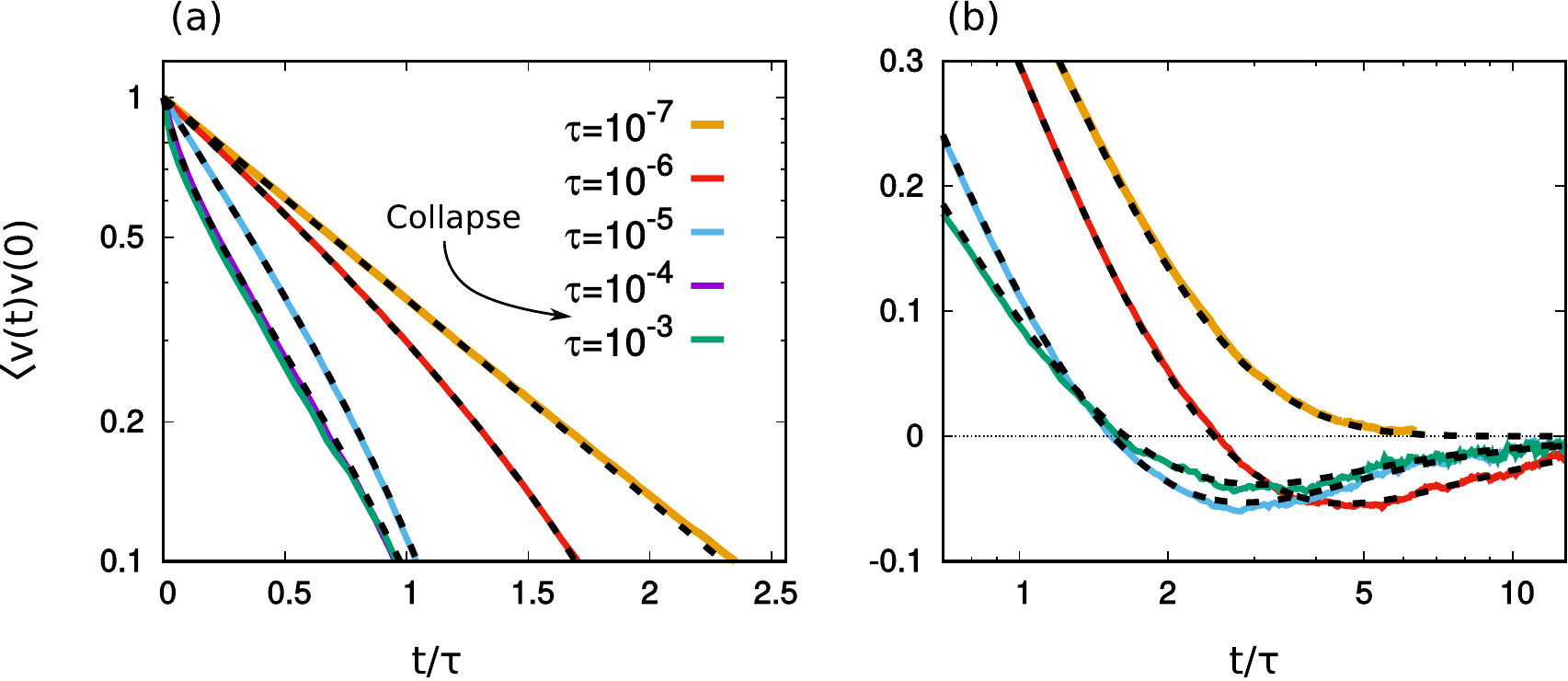}
\caption{\label{fig:autocorrelation} 
Velocity autocorrelation function, $\langle v(t)v(0)\rangle/\langle v^2\rangle$, as a function of $t/\tau$ for different values of $\tau$ as shown in the legend. Panels (a) and (b), correspond to different magnifications of the same observable.
Colored solid lines are obtained from numerical simulations while dashed black lines 
are the theoretical predictions of Eq.~\eqref{eq:theoretical_prediction_time}.
The horizontal dotted black line indicates the zero to provide an eye-guide.
Simulations are realised with $v_0=50$ and the interaction is given by Eq.~\eqref{eq:Ulj}.
}
\end{figure*}

In the previous section, we have investigated numerically and theoretically the spontaneous formation of velocity domains 
by studying the velocity correlation functions. We have seen that, despite the absence of any alignment interactions, 
the velocities of different particles develop a correlation which increases with the the persistence of the active force. 
%become spatially correlated over distances up to the correlation length.
In this section, we investigate the time-dependent properties of the system, considering, in particular,
the velocity autocorrelation and the two-time spatial velocity correlation function to unveil the dynamics of the velocity domains and estimate their life-times and permanence.

\subsection{Velocity autocorrelation function }

In Fig.~\ref{fig:autocorrelation}, we report the steady-state normalised  velocity autocorrelation functions (VACF), $\langle v(t) v(0) \rangle /\langle v^2 \rangle$,  as a function of $t/\tau$ for several values of $\tau$, and explore both the large and the small persistence regimes where velocities are spatially uncorrelated.
As shown in Fig.~\ref{fig:autocorrelation}~(a), this observable decays within a typical time of order $\tau$ for the whole range of persistence times numerically explored. 
In the  small $\tau$ regime,
i.e. when $\tau U''(\bar{x})/\gamma \ll  1$, the VACF displays the same exponential shape as the active-force autocorrelation, i.e.
$\langle \text{f}^a_i(t) \text{f}^a_i(t')\rangle= \frac{D\gamma^2}{\tau} e^{-|t-t'|/\tau}$,  (see the yellow curves in Fig.~\ref{fig:autocorrelation}).
Such a behaviour is a direct consequence of Eq. \eqref{eq:velocity_dynamics} when $\tau U''(\bar x)/\gamma\ll 1$
 since, in this limit, $\Gamma_{ij}\approx\delta_{ij}$.
When $\tau$ increases, the shape of the VACF changes and the relaxation process  becomes faster as measured 
with respect to rescaled time $t/\tau$.
Then, for further values of $\tau$, in particular, for $\tau \gtrsim 10^{-4}$, the rescaled VACFs collapse onto the same curve as shown in Fig.~\ref{fig:autocorrelation}.

Quite surprisingly, the VACFs $\langle v(t) v(0) \rangle /\langle v^2 \rangle$ assume negative values for $t>\tau$, except in the small-$\tau$ regime, for $\tau U''(\bar{x})/\gamma \ll 1$, where the decay is exponential as already discussed. 
In the former case, the autocorrelations decay very slowly (with a power-law behaviour) towards zero from negative values, as zoomed in Fig.~\ref{fig:autocorrelation}~(b).
The sign inversion of $\langle v(t) v(0) \rangle /\langle v^2 \rangle$ is understood in terms of the presence of two
competing mechanisms, the active force producing a negative contribution to the correlations and the restoring force of the individual oscillation modes driving the VCF towards its final value. 
Each mechanism is characterised by a different time scale, $\tau$ and $\gamma/\omega_q^2$, respectively. 
Only if the active force acts on a time-scale not too small compared with $\gamma/\omega_q^2$
the above phenomenon can be observed, therefore it has not a passive Brownian counterpart and represents a pure non-equilibrium collective effect.

\subsection{Spatio-temporal velocity correlation functions}

%%%%%%%%%%%%%%%%%%%%%%%%%%%%%%%%%%%%%%%%%%%%%%%%%%%%%%%%%%%%%%%%%%%%%%%%
%----------------------------FIG.6---------------------------------------
\begin{figure*}[t!]
\includegraphics[clip=true,keepaspectratio,width=0.98\textwidth]
{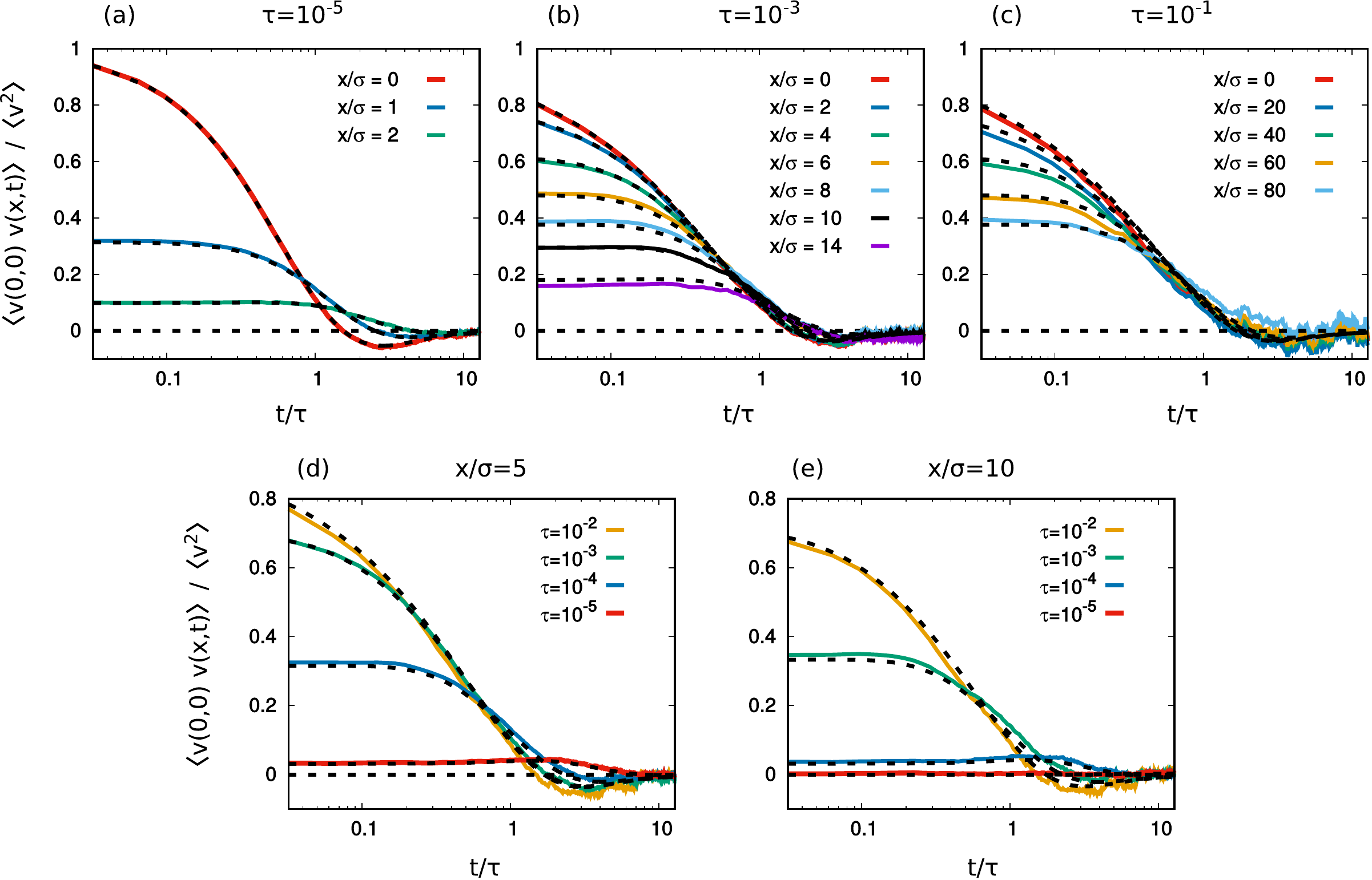}
\caption{\label{fig:spatiotemporalcorrelation} 
Spatio-temporal correlations of the velocities, $\langle v(x, t)v(0, 0)\rangle/\langle v^2\rangle$ as a function of $t/\tau$.
Panels (a), (b) and (c) correspond to $\tau=10^{-5}, 10^{-3}, 10^{-1}$, respectively. Each curve is obtained at different distances $x/\sigma$, as reported in the legends. Instead,
panels (d) and (e) show $\langle v(x, t)v(0, 0)\rangle/\langle v^2\rangle$ at $x/\sigma=5, 10$, respectively. As indicated in their legends, each curve is realized with a different values of $\tau$.
Colored solid lines report numerical simulations data while dashed black lines 
are predictions according to Eq.~\eqref{eq:theoretical_prediction_time}.
The lower dashed black line indicates the zero and corresponds to the spatio-temporal correlations for $x\gg \ell$. 
Simulations are realised with $v_0=50$ and the interaction is given by Eq.~\eqref{eq:Ulj}.
}
\end{figure*}
%------------------------------------------------------------------------

Fig.~\ref{fig:spatiotemporalcorrelation}~(a)-(c) displays the spatio-temporal velocity correlation function (VCF), $\langle v(x, t) v(0, 0) \rangle /\langle v^2 \rangle$, as a function of $t/\tau$ for several distances, $x/\sigma$, roughly from $x=0$ to $x\sim \ell$. 
In particular, panel (a), (b) and (c) are obtained at different values of $\tau$, corresponding to configurations ranging
from those with negligible  spatial correlations to those highly correlated.
For the smallest value of $\tau$, reported in panel (a), even nearest-neighbour pairs sitting at distance $x/\sigma=1$ are practically uncorrelated in time, while for larger $\tau$-values very far particles display a pronounced time-correlation (panels (b) and (c)).
In the cases (b) and (c), the spatio-temporal correlations  for ``small''  separation $x/\ell$ closely resemble the corresponding VACF.
On the contrary, for $x \sim \ell$, the VCF  $\langle v(x, t) v(0, 0) \rangle /\langle v^2 \rangle$ displays a sort of plateau for an initial time-window (always $\lesssim \tau$) until it collapses onto the VACF curve for $t \gtrsim \tau$. 

At the origin of such a plateau is the fact that the state of the particle placed at $x=0$ changes under the influence of the active force only after a time of order $\tau$. However,
the information about such a change does not reach a second particle belonging to the same velocity domain (hence correlated with the first) and
located $n$-sites away, before a time which increases with their separation has elapsed.
Finally, for $x\gg \ell$, any sort of spatio-temporal correlation is absent because the two particles do not belong to the same domain.

Based on the observation of the collapse of the VCF data onto the VACF curve for $x\sim\ell$, we identify the average life-time, $t^*$, of a domain with typical size $\ell$ as the time at which the normalised autocorrelation approaches the value $1/e$ being $e$ the Neper number.
This typical time $t^*$ shows a linear increase with $\tau$ above the dashed black line in Fig.~\ref{fig:lifetimes} which marks the first value with non-vanishing spatial correlations.
Below this line, even nearest neighbour particles are almost independent and velocity domains do not form.
In this last case, since the domain contains only a single particle, $t^*$ is nothing but the VACF relaxation time.
Such a quantity coincides with $\tau$ in the equilibrium-like regime where $\tau U''(\bar{x})/\gamma \ll 1$ and shows a non-linear increase as a function of $\tau$ in the crossover regime before the linear for $\tau U''(\bar{x})/\gamma \gg 1$.
%growth in the large persistence regime.

\subsection{Theoretical predictions}

To predict the shape of the spatio-temporal VCF, we consider the correlation of th Fourier modes as in the case of the steady-state spatial velocity correlations.
Solving the dynamics in the active harmonic crystal approximation, we obtain:
\begin{equation}
\label{eq:theoretical_prediction_time}
\langle \hat v_q(t) \hat v_{-q}(0)  \rangle= 
  v_0^2  \frac{ \Bigl[\frac{1}{\tau}  e^{-\frac{1}{\tau} t}-\frac{\omega^2_q}{\gamma} e^{- \frac{\omega^2_q}{\gamma} t }   )\Bigr]  
}{\left(1+\frac{\tau}{\gamma}\omega_q^2 \right) \left( \frac{1}{\tau}-\frac{\omega^2_q}{\gamma}\right)} \,,
\end{equation}
where the factor $\omega_q^2$ is defined by Eq.~\eqref{eq:def_omega2}.
At variance with the spatial velocity correlations (Eq.~\eqref{eq:vvprediction_Fourierspace}), here, the approximation for small $q$ is no longer valid to predict the whole time-behaviour (see Appendix~\ref{app:spatio_temporal} for more details about the derivation and the approximations involved).
The reason is that the wave-vector dependent VCF represented in Eq.~\eqref{eq:theoretical_prediction_time} is made of two contributions
each varying with its own relaxation time and both containing a divergence. Only by handling them together the two divergences cancel out. 
%As a consequence, to get the explicit space-time correlations, the integrals must be evaluated jointly.
Both for VACF and spatio-temporal VCF, Figs.~\ref{fig:autocorrelation} and~\ref{fig:spatiotemporalcorrelation} show the excellent agreement between data obtained via numerical simulations and the numerical integration of Eq.~\eqref{eq:theoretical_prediction_time} (normalised with $\langle v^2 \rangle$).

\subsubsection{Short-time approximation}

%%%%%%%%%%%%%%%%%%%%%%%%%%%%%%%%%%%%%%%%%%%%%%%%%%%%%%%%%%%%%%%%%%%%%%%%
%----------------------------FIG.7---------------------------------------
\begin{figure}[t!]
\includegraphics[clip=true,keepaspectratio,width=0.98\columnwidth]
{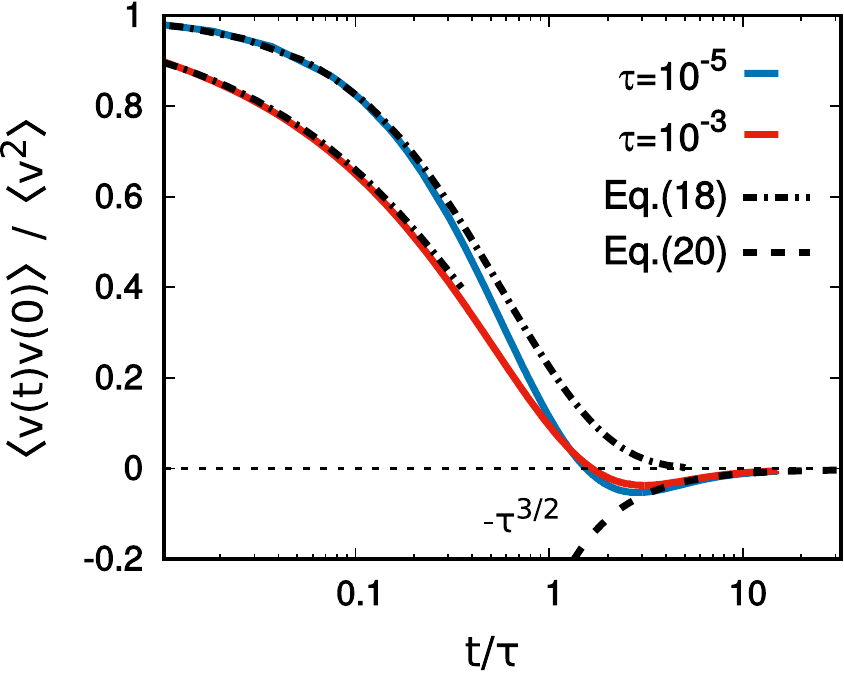}
\caption{\label{fig:pred} 
Velocity Autocorrelation, $\langle v(x, t)v(0, 0)\rangle/\langle v^2\rangle$, for two different values of $\tau$ as shown in the legend. The dashed black line is obtained by Eq.~\eqref{eq:approximate_largetimes}, while the two dashed dot black lines are obtained using Eq.~\eqref{eq:approx_smalltimes}.
}
\end{figure}
%------------------------------------------------------------------------

The integral in Eq.~\eqref{eq:theoretical_prediction_time} can be evaluated numerically in a straightforward way
but must be handled with care to extract analytical predictions about
the temporal decay of the spatial autocorrelation function. 
In the Appendix~\ref{app:spatio_temporal}, we derive a suitable approximation, holding for $t\lesssim \tau$, which consists in expanding the exponentials for small $(t/\tau)$ and resumming a class of terms.
In this way, the small-time decay of the autocorrelation is predicted and reads:
%predicts that the leading term in the correlation is
\begin{flalign}
\label{eq:approx_smalltimes}
\langle v(x, t) v(0,t)\rangle \approx v_0^2 e^{-t/\tau} \Omega\left(x, t \right) \,,
\end{flalign}
where the integral is given by
\begin{equation}
\Omega\left(x, t \right)= \int_{-\pi}^{\pi} \frac{dq}{2\pi} \,  \cos\left(q \frac{x}{\bar{x}}\right) \frac{e^{- 2(1-\cos(q)) U''(\bar{x}) t /\gamma}}{1+2(1-\cos(q))\frac{\tau}{\gamma} U''(\bar{x})} \,.
\end{equation}
This class of integrals can be performed exactly in terms of modified Bessel functions of the first kind and integer order~\cite{abramovitz1964handbook}.
Prediction~\eqref{eq:approx_smalltimes} fairly agrees with data in a time-window smaller than $\tau$ while the differences between small-time theory and simulations occur approximatively for $t>0.3 \,\tau$, as shown in Fig.~\ref{fig:pred} for two different values of $\tau$.
In particular, Eq.~\eqref{eq:approx_smalltimes} cannot reproduce the negative values assumed by the autocorrelation for $t\sim\tau$.
 In addition, in the short-time regime $t/\tau<1$, and, in particular, from Eq.~\eqref{eq:approx_smalltimes},
 it is possible to derive the leading correction of the departure of $\langle v(x, t) v(0, 0) \rangle$ from its equal-time value:
\begin{equation}
\label{eq:prediction_plateau}
\langle v(x, t) v(0, 0) \rangle-\langle v(x, 0) v(0, 0) \rangle \approx -
\frac{v_0^2}{(n+1)!}\frac{t}{\tau} \Bigl( \frac{t}{2\tau_K} \Bigr)^n  \,,
 \end{equation}
where $\tau_K=\gamma/2U''(\bar{x})$ and $n\approx x/\bar{x}$.
This prediction explains the plateau observed for large distances when $t/\tau$ is small.
Indeed, the prefactor of the first-order time-correction in Eq.~\eqref{eq:prediction_plateau} becomes smaller as $x$ increases through the $n$-dependence. 
In agreement with the numerical observations, a longer time is needed before the VCF with large $x$ deviates from $\langle v(x, 0) v(0, 0) \rangle$.

%{\color{red} Prova stirling per approssimare il fattoriale e passare ad una descrizione continua}

\subsubsection{Long-time behaviour}

As already mentioned, in the small persistence time regime %(i.e. $\tau U''(\bar{x}) \ll \gamma$)
the VACF has a pure exponential decay and, thus, for $t\gtrsim\tau$, approaches zero from positive values.
As numerically revealed, the decay is more complex in the large persistence regime, i.e. $\tau U''(\bar{x})/\gamma \gtrsim 1$.
While the VACF and the VCF are positive in the early stage of the relaxation process as shown by direct inspection of Eq.~\eqref{eq:theoretical_prediction_time}, they become negative at 
larger times. 
Each mode gives a positive contribution to the $q$-integral up to a wave-dependent crossover time, $t^*_q=-\tau
\ln (\tau \omega_q^2/ \gamma) /(1-\tau \omega_q^2/ \gamma)>0$, 
defined as the instant when the function in Eq.~\eqref{eq:theoretical_prediction_time} shows its first zero.
Such a $q$-dependent time decreases as $q$ grows and, thus, the final decay towards zero of the VACF (for $t\gtrsim\tau$) is controlled by the relaxation of the long wave-length modes since the short wave-lengths give negligible contributions to the integral.
Roughly for $t>\tau$ the $e^{-t/\tau}$ term in formula~\eqref{eq:theoretical_prediction_time} plays a negligible role.
Performing the $q$-integration of Eq.~\eqref{eq:theoretical_prediction_time} 
we obtain the following long-time approximation: % according to Eq.~\eqref{tvelocorr}:
%the final expression:
\begin{eqnarray}
\label{eq:approximate_largetimes}&&
\langle v(x,t) v(0,0)\rangle \approx
% -v_0^2 \tau \int_{\pi}^{\pi} \frac{dq}{2\pi} \frac{1}{1+\frac{\tau}{\gamma}\omega_q^2}  \frac{\omega_q^2}{\gamma} e^{-\frac{\omega_q^2}{\gamma} t}\nonumber\\&& \approx
-\frac{v_0^2\tau } {2\sqrt{K \pi/\gamma}} t^{-3/2} \,,
\end{eqnarray}
 which shows that, as $t\to\infty$, the VACF vanishes from negative values with an inverse power-law behaviour, explaining the behavior observed in Fig.~\ref{fig:autocorrelation}.
The prediction~\eqref{eq:approximate_largetimes} is derived in the Appendix~\ref{app:spatio_temporal} and numerically checked in Fig.~\ref{fig:pred} for two different values of $\tau$ showing a good agreement with data.
In addition, we remark that, in the small persistence regime, the relevant amplitudes in Eq.~\eqref{eq:theoretical_prediction_time} remain positive, as expected for passive systems. 
Finally, it is interesting to realise that the non-monotonic behaviour of the two-time correlation function   
discussed above is different from the monotonic behaviour of the linear response function, i.e. 
the response of the system to a
a small impulsive perturbation on the positions of the self-propelled particles. 
The calculation, reported in Appendix~\ref{appendix_response},
clearly shows that response and correlation are not proportional.

%%%%%%%%%%%%%%%%%%%%%%%%%%%%%%%%%%%%%%%%%%%%%%%%%%%%%%%%%%%%%%%%%%%%%%%%%%%%%
\section{Conclusions}
\label{SecConclusions}
%%%%%%%%%%%%%%%%%%%%%%%%%%%%%%%%%%%%%%%%%%%%%%%%%%%%%%%%%%%%%%%%%%%%%%%%%%%%%
The present work contains results concerning both the steady-state and time-dependent spatial properties of 
a system of interacting active particles. To the best of our knowledge, such an analysis has not been performed so far, in particular, for what concerns the characterisation of the two-time correlations in the non-equilibrium steady state of self-propelled particles.

The interest in such a study is twofold: the ensemble averages of the single time observables are constant in time 
but already contain interesting information about the spatial organisation of the system, namely display the presence of 
domains where the velocities of the particles are strongly correlated.
The two-time observables reveal how the system responds to external stimuli or how a 
given spatial structure lasts in time.
Our study
employs a one-dimensional model of interacting self-propelled particles evolving under the Active Ornstein-Uhlenbeck 
dynamics, as a test system. The numerical study is conducted by choosing a truncated and shifted Lennard-Jones (LJ) potential and large packing conditions. 
As a preliminary step, we have studied the stationary properties of the system to characterise the size of the velocity domains and focused on the spatial velocity correlations. 
Then, we investigated the dynamics of the velocity domains studying the autocorrelation and the spatio-temporal velocity correlation functions and determined the average life-time of a typical velocity domain. 
Our numerical study has been supported by theoretical arguments to derive the temporal dependence of the correlations.
Thanks to the lack of an appreciable density of defects or voids we can faithfully describe the system 
as a one-dimensional lattice of coupled harmonic oscillators, whose elastic and spacing constants are derived from 
expanding quadratically the LJ  potential near its equilibrium values. The resulting active harmonic lattice 
can be solved analytically in terms of Fourier normal modes giving  explicit formulae for one-time and two-time observables.

The theoretical method employed in this paper bears some similarities with the one
used in active polymer theory~\cite{Kaiser2015,Winkler2017,natali2020local,osmanovic2017dynamics, Eisenstecken2016}, i.e. the shape of the excitation spectrum to make an example. % $\omega_q^2$ is the same for instance.
However, the observables, here considered, are quite different from those usually studied in the context of active polymers, described by one-dimensional chains in two or three-dimensional spaces. Those studies are mainly concerned with diffusive ~\cite{anand2020conformation}, or configurational properties~\cite{bianco2018globulelike, liu2019configuration,martin2020hydrodynamics}, while
the present work regards the dynamical properties of a one-dimensional single file system where the interplay between low dimensionality, steric interactions, and self-propulsion determines an off-equilibrium, highly correlated motion not studied so far. 

%%%%%%%%%%%%%%%%%%%%%%%%%%%%%%%%%%%%%%%%%%%%%%%%%%%%%%%%%%%%%%%%%%%%%%%%
\appendix
\section{The harmonic active crystal \label{SecBrown}}

In this Appendix, we illustrate the approximations employed to develop the theoretical formulas that in the main text 
 have been compared to the numerical data.
We replace the full LJ potential by its Taylor expansion around the average particles' positions, $\bar{x}$.
The expansion is truncated at the second order and gives:
\begin{equation}
U_{hc}=U''(\bar x) \sum_{i}^N (x_{i+1}-x_i-\bar{x})^2 \,.
\end{equation}
Replacing $U_{tot}$ with $U_{hc}$, the matrix $\Gamma_{ij}$ assumes a simple form and the dynamics~\eqref{eq:velocity_dynamics} becomes:
\begin{equation}
\begin{aligned}
\label{eq:app_vicsek_dynamics}
\tau\gamma\dot{v}_i =& -\gamma\left[ v_i+ \frac{\tau}{\gamma}\sum_{j=i\pm 1} U''(\bar{x}) (v_i - v_j) \right] \\
&+F_i+ \gamma v_0 \sqrt{2 \tau}  \xi_i  \,.
%&-2\tau U''(\bar{x}) \left( v_i - w_i \right) + F_i\,,
\end{aligned}
\end{equation}
Remarkably, the force term, maintains the average distance between neighboring particles and is derived from $U_{hc}$.
Nevertheless, its shape is not particularly significant.
Splitting the square brackets in Eq.~\eqref{eq:app_vicsek_dynamics}, we obtain the dynamics~\eqref{eq:vicsek_dynamics}.
Representing Eq.~\eqref{eq:app_vicsek_dynamics}, in Fourier Space, we obtain
the equations of motion of the harmonic crystal:
\begin{eqnarray}
\label{eq:app_qdyn}
&&
\frac{d}{dt}\hat{u}_q(t) =\hat v_q \\&&
\frac{d}{dt}\hat v_q(t) =-\frac{\omega^2_q}{\gamma} \hat v_q(t)  -\frac{1}{\tau}  \hat v_q(t)
-\frac{\omega_q^2}{\gamma\tau} \hat u_q+v_0\sqrt{\frac{ 2}{\tau}} \hat \xi_q \nonumber
\end{eqnarray}
where $\hat{u}_q$ and $\hat{v}_q$ are the Fourier transforms of the displacement and velocity, respectively, which explicitly are
\begin{equation}
\left(
\begin{array}{c}
\hat u_{q}\\
\hat v_{q}\\
\end{array}
\right)=\frac{1}{\sqrt N}\sum_{n=1}^N \cos{q n}  
\left(      
\begin{array}{c}
(x_n-{n \bar x})\\
v_n\\
\end{array}
\right) \,,
\end{equation}
while   $\hat{\xi}_q$ is a noise term obtained from the fourier transform of $\xi_i$.
The reciprocal lattice  wavevectors are $q=2\pi k/N$ 
with $k=1,N$, while the frequency $\omega_q$ is such that: 
\begin{equation}
\label{eq:app_def_omega2}
\omega^2_q = 2 U''(\bar{x}) \left(1- \cos{\left(q \right)} \right) \,.
\end{equation}
We remark that, in the
limit $N\gg1$, the sums can be approximated by integrals
$$
\frac{1}{N}\sum_{n=1}^{N}  \to \frac{2}{\pi}\int_{\pi/N}^{\pi/2} dq \,.
$$

 In the following, we shall maintain a discrete notation that identifies each particle with an integer index.
The comparison between the  discrete theoretical and the $x$-dependent numerical observables is made by noting that, in the high-density regime considered, the particle's coordinates approximately satisfy the relation $x=n\bar{x}$, 
making possible replacing $v_n$ by   $v(x)$.

%%%%%%%%%%%%
\section{Variance of positions and velocities}\label{app:variances}
Without further approximations, we can derive the equal time averages over the active noise fluctuations, calculating the steady-state time correlations of Eq.~\eqref{eq:app_qdyn}:
 \begin{eqnarray}
 &&
 \langle  \hat{u}_q(t) \hat{u}_{-q}(t)  \rangle= 
 \frac{\gamma}{\omega_q^2}  \frac{\tau v_0^2}{1+\frac{\tau}{\gamma}\omega_q^2}  \,,
   \\&&
\langle \hat{v}_q(t)\hat{v}_{-q}(t)  \rangle=
  \frac{v_0^2}{1+\frac{\tau}{\gamma}\omega_q^2}  \,.
\end{eqnarray}
Setting $\langle u^2\rangle=\langle (x_n-n\bar x)^2\rangle$ (which is $n$-independent) and $K=U''(\bar x)$, for the sake of simplicity, the resulting  autocorrelation of the displacement, $u$, reads:
%(we set $K=U''(\bar x)$ and $\langle u_n^2\rangle= \langle (x_n-n\bar x)^2\rangle $)
\begin{equation}
\label{eq:app_u2}
\langle u^2\rangle
=
\tau\frac{v_0^2 \gamma}{2\pi}\int_{\pi/N}^{\pi/2} dq \left( \frac{1}{4 K \sin^2( \frac{q}{2}  )}-
 \frac{\frac{\tau}{\gamma}}{1+\frac{\tau}{\gamma}4 K \sin^2( \frac{q}{2}  )} 
\right) \,.
\end{equation}
 The first integral in Eq.~\eqref{eq:app_u2} is divergent for $N\to\infty$, i.e. when performed from $0$ to $\pi/2$, at variance with the second one.
However, for a finite but large sample with $N\gg1$, by discarding a small interval near the origin we obtain from Eq.~\eqref{eq:variance_position} the following result % for arbitrary $\tau$ and $K$ for the displacement variance\\
\begin{equation}
\langle u^2\rangle=\tau\frac{v_0^2\gamma}{K}   \frac{1}{2\pi^2} N-   \frac{v_0^2\tau^2}{\left(1+4 \frac{\tau}{\gamma} K \right)^{1/2}} \,.
\label{varianceu}
\end{equation}
 Eq.~\eqref{varianceu} makes sense just if $\langle u^2\rangle>0$, meaning that does not hold for arbitrary values of $\tau$. Indeed, the validity of Eq.~\eqref{varianceu} is restricted to the regimes where the persistence length, $v_0 \tau$, is smaller than $L=N\bar{x}$ (i.e. the regimes numerically explored in this paper).
A simple integration gives  the formula for the velocity variance reported in Eq.~\eqref{eq:variance_velocity}.
Interestingly, we can relate the velocity fluctuation to the positional fluctuation through the following relation:
\begin{equation}
\label{unvar}
\langle u^2\rangle= \langle u^2\rangle_{\tau=0}-
\tau^2 \langle v^2\rangle \,,
\end{equation}
where the first term in the right hand side is the well known variance of the displacement fluctuations in the white noise limit, $\tau=0$~\cite{PhysRevLett.40.1507,yoshida1981dynamic}. 
Thus, the single units, comprising the active harmonic crystal, 
have fluctuations smaller than those of the corresponding \emph{passive} crystal. This observation agrees with the well-known results of single active harmonic oscillators, which display an enhanced rigidity with respect to their equilibrium counterparts.

%%%%%%%%%%%%%%%%%%%%%%%%%%%%%%%%%%%%%%%%%%%%%%%%%%%%%%%%%%%%%%%%%%%%%%%%

\section{Spatial correlations of the velocity and energies}\label{app:SpatialCorrealation}
The equal-time velocity pair correlation function in Eq.~\eqref{eq:spatial_corr_theory} is obtained by expanding 
$\omega_q^2$ up to quadratic order in $q^2$ and extending the Fourier integrals to the whole real axis. 
Such an approximation is valid when the discretization of the lattice becomes irrelevant.
Let us write
\begin{eqnarray}&&
\label{eq:app_spatial_vel_cor}
\langle  v_{i+n}(t) v_i(t)  \rangle= \frac{v_0^2}{ 2\pi}\int_{-\pi}^{\pi} \, dq
  \frac{\cos(nq)}{1+2\ell^2 (1-\cos(q))}  \nonumber\\&&
 =  \frac{v_0^2}{2 \ell^2}\, g_n(\frac{1+2\ell^2}{2\ell^2})
    \end{eqnarray}
where $g_n(x)=\frac{1}{2\pi}   \int^{\pi}_{-\pi}    dq\,   \frac{\cos(n q) }{x-\cos(q)}$.
For small values of the index $n$, $g_n$ can be expressed in terms of elementary functions,
for instance $g_1(x)= \, \left[ \frac{x}{\sqrt{x^2-1}} -1\right]$,
  $g_2(x)= \left[ \frac{2x^2-1}{\sqrt{x^2-1}} -2 x\right]
$, $g_3(x)= \left[ \frac{4x^3-3x}{\sqrt{x^2-1}} -(4x^2-1) \right]$, whereas in general
we have to write it in terms of Hypergeometric functions. By expanding the result in powers of $\ell$,
one verifies that the terms are very well reproduced by the formula~\eqref{eq:spatial_corr_theory}.
 This procedure yields the same result as if we expand up to the second order in $q$ 
 the denominator in Eq.~\eqref{eq:app_spatial_vel_cor} and perform an integration over the whole real axis.

By a simple extension of the previous method we may evaluate the energy correlation function. In our discrete notation
we define $E_l=v_l^2$ and consider
 \begin{equation}
\langle (E_{l+m}-\langle E_{l+m} \rangle )  ( E_l -\langle E_l \rangle)      \rangle=
2\langle v_l v_{l+m} \rangle \langle v_l v_{l+m}\rangle
 \end{equation}
where we have used the Gaussian property
of the distribution to factorize the operator average. Taking into account Eq.~\eqref{eq:spatial_corr_theory}, we obtain the 
final expression:
\begin{equation}
\langle  (E_{l+m}-\langle E_{l+m} )( E_l -\langle E_l \rangle) \rangle )=
 2 \langle v^2\rangle^2  e^{- 2 |m|\bar x/\ell} \,,
  \end{equation}
which coincides with Eq.~\eqref{eq:spatialenergy_corr}.

%%%%%%%%%%%%%%%%%%%%%%%
\section{Spatio-temporal correlation functions}\label{app:spatio_temporal}
For $t>t'$, the Fourier Transform of the two-time displacement correlation function, in the steady-state, reads
\begin{equation}
\langle \hat u_q(t) \hat u_{-q}(t')  \rangle= v_0^2\frac{\tau\gamma}{  \omega^2_q}
 \frac{\Bigl[\frac{\omega^2_q}{\gamma}e^{-\frac{1}{\tau} (t-t')   } -\frac{1}{\tau}  e^{- \frac{\omega^2_q}{\gamma} (t-t') }  )\Bigr] }{(\frac{\omega^2_q}{\gamma}-\frac{1}{\tau})\left(1+\frac{\tau}{\gamma}\omega_q^2\right)} \,,
    \end{equation}
 while the Fourier Transform of the  velocity correlation function is given by Eq.~\eqref{eq:theoretical_prediction_time}.
 Thus, to come back to real space, we have to calculate the integral
\begin{equation}
\label{velvel}
\langle v _n(t) v_0(0)\rangle =  \frac{v_0^2}{\pi}\int_0^{\pi} dq 
\frac{ \cos(q n)}{1+\frac{\tau}{\gamma}\omega_q^2}  \frac{ \Bigl[\frac{\omega^2_q}{\gamma} e^{- \frac{\omega^2_q}{\gamma} t } -\frac{1}{\tau}  e^{-\frac{1}{\tau} t}  )\Bigr] }{(\frac{\omega^2_q}{\gamma}-\frac{1}{\tau})} \,.
    \end{equation}
The apparent singularity of the denominator is eliminated by the concomitant vanishing of the numerator.

\subsection{Short-time approximation in real space}
Let us consider the following integral
\begin{equation}
\frac{d}{dt} \sigma^2_n(t)=
\frac{v_0^2 \tau }{\pi}  \int_{-\pi}^\pi dq \,\frac{ \cos(qn)}{1+\tau\frac{\omega_q^2}{\gamma}    }\, \frac{\left[e^{ - \tau \frac{\omega^2q}{\gamma}   } -e^{-\frac{1}{\tau} t}  \right]}{ 1-\tau\frac{\omega_q^2}{\gamma}}
\label{dsndt}
   \end{equation}
related to the velocity correlation by the identity $\langle v _n(t) v_0(0)\rangle =\frac{d^2 \sigma^2_n(t) }{dt^2} $.
After Taylor expanding the first and second exponential in powers of the rescaled time, $t/\tau$, we rewrite it as
\begin{eqnarray}
&&\frac{d}{dt} \sigma^2_n(t)=-\frac{v_0^2 \tau }{\pi}
 \int_{-\pi}^\pi dq \,\frac{ \cos(qn)}{1+\tau\frac{\omega_q^2}{\gamma}} \nonumber\\&&
\times \frac{1}{ 1-\tau\frac{\omega_q^2}{\gamma}}\left[ \frac{t}{\tau}(1-\tau\frac{\omega_q^2}{\gamma}) - \frac{t^2}{2\tau^2}\left(1-(\tau\frac{\omega_q^2}{\gamma})^2\right) \right]+O( \frac{t^3}{\tau^3})\,. \nonumber
\end{eqnarray}
Up to this order, we may approximate the above result by the following expression:
\begin{equation}
\frac{d}{dt} \sigma^2_n(t)
=-\frac{v_0^2 \tau }{\pi}  \,    \int_{-\pi}^\pi dq \,\frac{ \cos(qn)}{(\Gamma(q))^2   } (1-e^{-\Gamma(q) t/\tau}) \,,
   \end{equation}
   with $\Gamma(q)=1+\tau\frac{\omega_q^2}{\gamma}$.
   Finally, differentiating with respect to $t$, we get the short-time formula for the velocity correlation function:
\begin{equation}
\langle v_n(t) v_0(0) \rangle= \frac{v_0^2}{ 2\pi}  \int_{-\pi}^\pi dq \, \frac{\cos(qn)}{\Gamma(q)} 
e^{-\Gamma(q) t/\tau} \,.
  \end{equation}
  Perhaps, the best strategy to display the presence of a small $t$ interval where the velocity two-time correlation function 
  for particles separated by a distance $n$ is to study the first time derivative of such a function:
\begin{equation}
\begin{aligned}
-\frac{d}{dt}\langle v_n(t) v_0(0) \rangle= &
\frac{1}{ 2\pi} \frac{v_0^2}{\tau} 
 e^{-t/\tau}e^{- \frac{t}{\tau_K}} \\
&\times\int_{-\pi}^\pi dq \, \cos(qn)  e^ { \frac{t}{\tau_K} \cos(q) } \,,
\end{aligned}
\end{equation}
where $\tau_K=\frac{\gamma}{2 K}$. The integration can now be performed exactly:
\begin{equation}
-\frac{d}{dt}\langle v_n(t) v_0(0) \rangle=  \frac{v_0^2}{\tau} 
 \frac{e^{-t/\tau}}{ 2\pi}e^{- \frac{t}{\tau_K}} \, I_n(\frac{t}{\tau_K}) \,,
 \end{equation}
where $I_n(x)$ is the first-kind modified Bessel function of order $n$.
Using standard properties of the Bessel functions, we find the following short-time aproximation:
\begin{equation}
\langle v_n(t) v_0(0) \rangle\approx\langle v_n(0) v_0(0) \rangle-
\frac{1}{n!}\frac{v_0^2}{\tau} \Bigl( \frac{1}{2\tau_K} \Bigr)^n \frac{ t^{n+1}}{n+1}\,.
 \end{equation}
That is to say, the relaxation rate of the pair correlation becomes smaller with increasing separation of the
the two particle considered since it is proportional to $(t/\tau)$ to a power equal to $(n+1)$. This formula coincides with
Eq.~\eqref{eq:prediction_plateau} and explains the formation of plateau regions.
These plateau become more and more evident when the particle's
separation increases.

\subsection{Long-time approximation}
We define the crossover time, $t^*_q$, as the time at which the VACF vanishes. 
From Eq.~\eqref{eq:theoretical_prediction_time}, it is straightforward to see that $t^*_q=-\tau
\ln (\tau \omega_q^2/ \gamma) /(1-\tau \omega_q^2/ \gamma)$.
For times larger than  $t^*_q$, 
it is safe to neglect the contribution of the $e^{-t/\tau}$ term in Eq.~\eqref{velvel}.
The modes with small values of  $q$ yield the main contributions to the integrals Eq.~\eqref{velvel}
and the result is negative because the constant term in the denominator is larger than the $q$-dependent term.
To evaluate Eq.~\eqref{velvel} in the long-time regime $t/\tau\gg 1$, we first consider
the following integral:
 \begin{eqnarray}
\frac{d}{dt} \sigma^2_n(t) &&\approx
\frac{v_0^2 \tau }{\pi} e^{-(2K/\gamma) t} \int_{-\pi}^\pi dq \, e^{ 2 \frac{K}{\gamma} \cos(q) t   } \cos(qn)  \nonumber\\
&&=2 v_0^2\tau e^{-(2K/\gamma) t}  I_n(\frac{2K}{\gamma}t)
\label{eq107}
\end{eqnarray}
and use the following large-$t$ approximation
$$
\frac{d}{dt} \sigma^2_n(t)\approx
\frac{v_0^2\tau } {\sqrt{K \pi/\gamma}} t^{-1/2}
(1-\frac{4 n^2-1}{16}\frac{1}{(K/\gamma) t }+\dots) \,.
\label{q75}
$$
Finally, the sought correlation is found by differentiating such a formula with respect to $t$:
\begin{equation*}
\label{tvelocorr}
\langle v_n(t) v_0(0) \rangle
\approx
-\frac{v_0^2\tau } {2\sqrt{K \pi/\gamma}} t^{-3/2} (1-\frac{3}{32}\frac{(4 n^2-1)}{(K/\gamma) t }+\dots) \,.
 \end{equation*}
This equation coincides with Eq.~\eqref{eq:approximate_largetimes} at the first order and the other terms of the expansion are negligible because of the large value of $K$.

\section{Response function}
\label{appendix_response}

The impulse response function for the AOUP was obtained by Szamel~\cite{szamel2014self} for the harmonic case
and by Caprini et al.~\cite{caprini2018linear}  in the general case.
In the present Appendix, we derive the response of the system
by adding a small impulsive force $\hat h_q(t)= \hat h_q^0 \delta(t)$ to the equation describing the evolution of the 
$q$-component of the displacement.

Following a standard procedure (see for instance, Ref~\cite{marconi2008fluctuation}) to compute the response to an impulsive perturbation acting on a particular $q$-mode we arrive at
 the following formula 
 \begin{equation}
\label{eq:linearResponse_V}
R_q(t)= - \langle \hat u_{q}(t) \frac{\partial}{\partial \hat u_q} \ln P_q(t=0)\rangle \,,
\end{equation}     
where $P_q$ is the known steady state distribution of the $q$ mode, explicitly given by
\begin{equation}
P_q\propto e^{- \frac{\left(1+\frac{\tau}{\gamma}\omega_q^2 \right)}{2D   } \left[
   \frac{\omega_q^2}{ \gamma}  \hat u_q\hat u_{-q}+ 
   \tau\left((\frac{\omega_q^2  }{\gamma}) \hat u_q-  \hat \eta_q\right)  
   \left((\frac{\omega_q^2  }{\gamma}) \hat u_{-q}-  \hat \eta_{-q}\right)   \right] } \,.
\end{equation}
The response function, defined as:
 \begin{equation}
 R_q(t)= \frac{ \langle \hat u^{(h)}_{q}(t) -\hat u_q(t)\rangle}{\hat u^{(h)}_{q}(0) -\hat u_q(0)} \,,
   \end{equation}
 turns out to be
$$
R_q(t)=e^{-(\omega_q^2/\gamma) t} \,,
$$
using Eq.~\eqref{eq:linearResponse_V} and the linearity of the system.
Remarkably, the response function only depends on the frequency $\omega_q$ but not on the persistence time, $\tau$. 

%%
%\subsection{Entropy Production}

%The entropy production rate, $\dot\sigma$, of the AOUP system can be calculated using a Fokker-Planck approach, i.e. generalizing the result of Maggi, Marconi, Puglisi, Sci. Rep. 2017, to the present case and using Einstein repeated indexes
%convention we write:
%\begin{equation}
%\label{eq:sigma2}
%\dot\sigma= \frac{1}{\tau} \left\langle   Tr\left[\Gamma\right]  \right\rangle - \frac{1}{D} \left\langle  v_k \Gamma_{km}\Gamma_{mj}v_j  \right\rangle \,.
%\end{equation}

%Now, using the following exact formula for the harmonic model  
%$$
%\langle v_k v_j  \rangle=\Gamma^{-1}_{kj}
% $$
% and the cyclic property of the trace we find

%\begin{equation}
%\label{eq:sigma2bb}
%\dot\sigma= \frac{1}{\tau} \left\langle   Tr\Gamma  \right\rangle - \frac{1}{\tau}
%\Gamma^{-1}_{kj}\,\Gamma_{km}\Gamma_{mj}=0
%\end{equation}
%In other words, the entropy production in the harmonic approximation vanishes.

%\bibliographystyle{mdpi}
\bibliographystyle{rsc} %the RSC's .bst file

\bibliography{1Dbib.bib}

%%%%%%%%%%%%%%%%%%%%%%%%%%%%%%%%%%%%%%%%%%
%% optional
%\sampleavailability{Samples of the compounds ...... are available from the authors.}

\end{document}